\documentclass[12pt]{aastex}
\newcommand{\msun}{\hbox{$M_\odot$}}
\newcommand{\lsun}{\hbox{$L_\odot$}}

\slugcomment{Version from \today}

\begin{document}


\title{SPECTRAL ENERGY DISTRIBUTIONS OF HIGH-MASS PROTOSTELLAR OBJECTS - EVIDENCE FOR HIGH ACCRETION
RATES}

\author{F. M. Fazal\altaffilmark{1,2,7}, T. K. Sridharan\altaffilmark{1,3,7}, K. Qiu\altaffilmark{1,4}, T.
Robitaille\altaffilmark{5,7}, B. Whitney\altaffilmark{6,7}, Q. Zhang\altaffilmark{1,7}}
\altaffiltext{1}{Harvard-Smithsonian Center for Astrophysics, 60 Garden Street, Cambridge, MA 02138}
\altaffiltext{2}{Dept. of Physics, Amherst College,
Amherst, MA 01002; work done under the NSF/SAO-REU summer intern program}
\altaffiltext{3}{contact author}
\altaffiltext{4}{Department of Astronomy, Nanjing University, Nanjing, China}
\altaffiltext{5}{Scottish Universities Physics Alliance, School of Physics and Astronomy, Univ. of St Andrews,
\altaffiltext{6}{Space Science Institute, 4750 Walnut Street Suite 205, Boulder, CO 80301}
North Haugh, KY16 9SS, St Andrews, UK}
\altaffiltext{7}{e-mails: ffazal08@amherst.edu, tksridha@cfa.harvard.edu, kqiu@cfa.harvard.edu, bwhitney@spacescience.org, tr9@st-andrews.ac.uk, qzhang@cfa.harvard.edu}


\begin{abstract}
The spectral energy distributions (SEDs), spanning the mid-infrared to millimeter
wavelengths, of a sample of 13 high-mass
protostellar objects (HMPOs) 
were studied using a large archive of 2-D
axisymmetric radiative transfer models. Measurements from the Spitzer GLIMPSE
and MIPSGAL surveys and the MSX survey were used in addition to our own
surveys at millimeter and submillimeter wavelengths to construct the SEDs,
which were then fit to the archive of models. These models assumed that stars
of all masses form via accretion and allowed us to make estimates for the
masses, luminosities and envelope accretion rates for the HMPOs. 
The models fit the observed SEDs well. The implied  envelope accretion rates 
are high,
$\approx 10^{-2.5} \msun/yr$, 
consistent with the accretion-based scenario of
massive star formation. With the fitted accretion rates and with  mass estimates of 
up to $\sim 20\ \msun$ for these objects, it appears plausible that stars with 
stellar masses $M_{\ast} > 20 \msun$ can form via accretion. 

\end{abstract}

\maketitle


\section{Introduction}

The process of massive star-formation is a subject of current debate (McKee \&
Ostriker, 2007; Zinnecker, 2007). While
scaled up versions of the standard accretion model and competitive
accretion models require high accretions
rates ($\sim 10^{-3} \msun/yr$), competing coalescence models need high
stellar densities ($\sim 10^8 /pc^3$). Presumably, the models are
applicable in 
different regimes, but it is unclear if there is a limit to the stellar mass 
in the accretion models.  The ubiquitous detection of
outflows with derived high accretion rates supports an accretion picture 
(Churchwell 1999, Henning et al 2000, Beuther et al 2002, Zhang et al 2005). 
Observations of spectral line infall signatures also 
indicate high infall rates (Zhang \& Ho, 1997, Keto, 2002, Fuller, Williams \&
Sridharan, 2005, Beltran et al 2006, Keto \& Wood, 2006, Zapata et al
2008).  However,
these results are subject to assumptions and questions. Therefore, other independent 
methods of deducing
accretion and determining the accretion rates are important. In this paper, 
we model the SEDs of a subset of
13 HMPOs and find results that are consistent with the accretion scenario for massive
star-formation.  The study takes advantage of (1) the Spitzer legacy surveys
GLIMPSE and MIPSGAL 
to enhance wavelength
coverage for the SEDs at high spatial resolutions compared to previous data
from IRAS and MSX surveys and (2) new modeling capabilities.
Thus, the current work is a significant improvement over 
previous 1-D modeling of the same objects (Williams, Fuller \&
Sridharan, 2005).

\section{Sample and Data Analysis} The starting point is the full sample of 69 
objects
presented in Sridharan et
al (2002) which was chosen by combining the Wood \& Churchwell IRAS colors, lack of
or weak cm-wavelength continuum emission, high luminosities, and CS
detections.  Images of the sample at 1.3mm and  450   \& 850 $\mu$m
wavelengths showed strong emission (Beuther et al 2002, Williams, Fuller \& Sridharan, 2004). 
Of the 69 fields, 52 had MIPSGAL data at both
24 and 70 $\mu$m of which 41 and 4 were saturated in the two bands,
respectively. Attempts 
to overcome saturation using simple
PSF fitting were abandoned due to poor fits. Presence of multiple objects also
often prevented reliable photometry. We finally chose a subset
of 13 objects, listed in Table 1, for which (1) reliable fluxes could be
obtained at 70 $\mu$m and atleast 2 IRAC bands from the GLIMPSE survey catalog
(Spring '05, highly reliable), and (2) the GLIMPSE catalog position and
the 1.2mm image peak were coincident.
The kinematic distances listed, from Sridharan et al (2002), used CS velocities.
For cases where distance ambiguity is resolved here, the discarded distance is
listed in parenthesis.

The fluxes at 24 and 70 $\mu$m were obtained by aperture photometry after
background subtraction using
the MOPEX/APEX package from the Spitzer Science Center (SSC). 
Fluxes for the IRAC and MSX bands were obtained from
the GLIMPSE Catalog and the MSXC6 Source Catalog
respectively. 
Aperture correction factors recommended by the SSC
were used for MIPS data. 
The 2MASS data are not included due to questionable associations because of the large and
uncertain extinctions.  
A lower limit of $25 \%$ error was imposed on the measurements to account for
uncertainties in the absolute flux calibration, photometry extraction
and the possibility of variability.  An upper limit ($90 \%$
confidence) was imposed on some of the fluxes, particularly in the MSX bands,
if the source had nearby luminous objects that were likely
to have contributed significantly ($>15-20\%$) to the flux.

Combining the Spitzer data with measurements at 1.3mm, 450 and \& 850 $\mu$m
and with MSX data (wherever GLIMPSE data were not available) allowed us to construct 
SEDs for the 13 objects over a wide wavelength range. 
We then searched for models that were best fits to the SEDs in a 
large archive of two-dimensional (2D) axisymmetric radiative 
transfer models of protostars
calculated for a large range of protostellar masses, accretion rates,
disk masses, and disk orientations.
(Robitaille et al 2006, 2007\footnote{available at
http://www.astro.wisc.edu/protostars}).
This archive has a linear regression tool that can select all model
SEDs that fit the observed SED better than a specified $\chi^2$.  Each
well-fit SED has a set of model parameters corresponding to it, such
as stellar mass, temperature and age, envelope accretion rate, disk
mass and envelope inner radius. The models assume that stars of all
masses form via free-fall rotational collapse to a disk and accretion
through the disk to the star (Ulrich 1976; Terebey, Shu, \& Cassen
1987; Whitney et al. 2003a,b); thus the envelope accretion rate is
one of the parameters that sets the envelope mass. This analysis
will therefore serve as a test of the accretion scenario for massive
star formation. The grid of models samples
masses and ages of 0.1 -- 50 \msun\ and  $10^3 - 10^7$ yrs respectively. The stellar luminosity
and temperature are related to the age and mass through 
evolutionary tracks of Bernasconi \& Maeder (1996).  Similar studies
as this exist in the literature (De Buizer et al. 2005, Shepherd et al
2007, Indebetouw et al 2007, Simon et al 2007, Kumar \& Grave 2007),
although they primarily covered lower mass ranges, did not
specifically pick the earliest stages of massive star formation, used
only a few colors, or did not model a sample of sources. \\

For each source, a range of visual extinction and distance uncertainty
were explored: 0--30 magnitude; $\pm$ 0.5 kpc for the near distance and 
$\pm$ 1 kpc for the far distance and for cases with unambiguous distances 
except for the near distance for IRAS 19266+1745, where a range of 
0.1 -- 0.8 kpc was used.  
For sources with
distance ambiguity, fits were made for both distances.
If more than one measurement was available at a
given wavelength, the highest quality data were used. Thus, the
following two rules were imposed: (1) when IRAC $8.0\ \mu
m$ data was available the MSX A band data were not used, and (2) when the
MIPS $24\ \mu m$ data was available the MSX E band was not used. 
Aperture radii used for the fits are 
9$''$ and 27$''$  for the two MIPS bands, 30$''$ for MSX data, 5$''$
for IRAC data and different values for sub-mm/mm data, based on measured source
sizes.
Figure 1 shows the 
SEDs of the best-fitting model for three representative cases \footnote{
SEDs for discarded distances not presented}, 
along with all models that
fit the data reasonably well - with  $\chi^2 - \chi_{best}^2 < \Delta
N$, where $\Delta N = 3$. This arbitrary cut off based on visual
examination of the fits was adopted to give a 
reasonable number of models and to be
consistent with that assumed by Robitaille et al (2007).
In cases where this yielded
less than 5 models, we increased $\Delta N$ to 5 (IRAS 19217+1651 and the
far distance cases of IRAS 18264--1152, IRAS 18372--0541, IRAS 18472--0022 and IRAS 18553+0414), 
to allow better estimates of
the parameters and their variances. For the near distance case of IRAS 18553\ + 0414,  $\Delta N
= 1$ was adopted to
restrict the number of models to 200.


\section{Results and Discussion}

From the fits shown in Figure 1 it is apparent that
the best-fitting models match the SED data quite well.
The poorest fit is for IRAS18264$-$1152 where higher resolution observations
show the presence of multiple objects (Qiu et al 2007). Unresolved
multiplicity is a limitation of the study and we take the quality of the
fits for the other sources as an indication 
that the SED and envelope structure are likely primarily affected
by the most massive star in the system. 
Our primary conclusion from these fits is that the models provide a good description of
the SEDs of the HMPOs.

The parameters obtained from the SED fitting, viz., the stellar mass, luminosity
\& temperature, $M_{\ast}$,
$L_{\ast}$ \& $T_{\ast}$, the envelope inner radius $R_{in}$ and the envelope accretion rate
$\dot M_{env}$ are listed in Table 2.
While a formal confidence interval measure would be
desirable, the model grid sampling a 14-dimensional space makes it
difficult to do so. Nonetheless average values and variances for the parameters were
obtained by taking a weighted average of all the models within a ${\chi}^2
- {\chi_{best}}^2 < \Delta N$, following Simon
et al (2007). For a majority of the sources $M_{\ast}$ was
found to be constrained to within 3 $\msun$ and $L_{\ast}$, $T_{\ast}$ and $\dot M_{env}$
to within  0.5, 0.7 and 0.5 orders of
magnitude. For the envelope inner radius $R_{in}$, which did not seem to be well
constrained,  a range of values is listed.

We restrict further discussion to 9 sources with no or resolved 
distance ambiguity.  IRAS 18440--0148, IRAS 19035+0641, IRAS 19074+0752 and IRAS 19217+1651 have unambiguous 
kinematic distances. 
For IRAS 18553+0414 and IRAS 19266+1745, the near distance was discarded
by Williams et al (2004) because of incompatible dust mass and luminosity estimates, which
we confirmed by the SED fits. For IRAS 18247--1147, IRAS 18372--0541 
and IRAS 18472--0022, we compared spectral type estimates by two different methods and 
picked
the distance that resulted in better match. Lyman continuum photon rate
estimates using 3.6 cm continuum flux data from Sridharan et al (2002) yielded
a spectral type (De Buizer et al 2005,  Panagia 1973; strictly, a lower
limit). The second estimate  came
from SED fits. IRAS 18372--0541 and IRAS 18472--0022 were placed at the
far distance with confidence, whereas for IRAS 18247--1147 the choice of near distance was favored,
although less certain.  
This analysis could not be extended to 
IRAS 18090\ -- 1832, IRAS 18264\ -- 1152 and IRAS 18521\ + 0134 because
the corresponding 3.6 cm continuum flux was either undetected or was $<1$
mJy (Sridharan et al 2002).
For IRAS 18431--0312, the near and far 
distances, which are within $\pm$ 10\%, led to the same best fit model mass, envelope accretion rate, 
luminosity and the temperature within the uncertainties. Hence these parameters were averaged
together.

The model masses, luminosities and temperatures are found to be spread over the ranges 
$\sim$ 10 -- 20 \msun, $10^{3.5}-10^{5} \lsun$ and $10^4 - 10^{4.5}$ K respectively.
Figure 2 shows the fitted accretion rates as a function of the stellar masses,
with the main conclusion that we have high accretion rates:  
$\approx 10^{-2.5} \msun/yr$. 
Current evidence for high accretion rates come from estimates
based on high molecular outflow rates and infall studies 
(see references in section 1).
These 
estimates are subject to a 
number of uncertainties - the velocities of the underlying jets, the projection 
angles, the fraction of the accreting material ejected in the jets, the size of
the infall region and densities.
Our result is an independent confirmation of high accretion rates using a different approach, 
and lends important credence to the accretion based massive star-formation 
scenario. We note that in our models, turbulence and magnetic
fields are not included which affect the density profiles in the 
outer regions leading to uncertainties in the accretion rates.

Since all the objects show high accretion rates, this phase is
not transient.  With model masses up to $\sim 20 \msun$,  and assuming the high
accretion rates derived to continue for $\sim$ 10$^3$ years, it appears plausible that stars of masses $>$
20 $\msun$ could form by accretion. 

In comparison, outflow and infall studies mentioned above arrived at $\dot M_{disk}$
of $\sim 10^{-3} - 10^{-4} \msun /yr$ and a similar range for the infall rate
$\dot M_{env}$.
$\dot M_{env}$ and $\dot M_{disk}$, are the rates of material falling 
on to the disk from the envelope, and the accretion rate from the circumstellar disk 
to the protostar respectively, a distinction often not carefully made in
the literature. In the absence of episodic phenomena, the difference between
the two represents the rate of flow in the outflow jets.

While our accretion rates appear to be consistent with being independent of mass, there 
may be a weak trend of increasing $\dot M_{env}$ with stellar mass $M_{\ast}$ 
(Fig 2). Although uncertain, we carry out a formal power law fit to allow comparison 
with evolutionary tracks, and obtain the best fit of 
$ \dot M_{env} = 10^{-4.2 \pm 1.1} \times {M_{\ast}}^{1.4 \pm 0.9}  \msun/yr$. 
Heeding Robitaille (2007), who pointed out several
caveats to SED modeling using this archive, 
we conducted checks to convince ourselves that the trend suggested
is not due to biases from the model grid.
The distribution of all the grid points
within the relevant range in the model archive
and its best fit, both presented in Fig 2, demonstrate this.
Guided by outflow results, Norberg and Maeder (2000) included mass dependent  
$\dot M_{disk}$ for massive stars in their models of evolutionary tracks and 
found that a power law $\dot M_{disk} = 10^{-5} \times {M_{\ast}}^{1.5} \msun/yr$  best agreed
with observations of Pre-Main Sequence (PMS) stars in the HR diagram. Our 
tentative result is similar to this. 
Since the evolutionary tracks of Bernasconi \& Maeder (1996), used in the 
SED modeling do not incorporate accretion (the so called canonical models) and our accretion rates come from
collapse models and density profiles, the agreement with
the later Norberg and Maeder (2000) work on evolutionary tracks with accretion 
points to the potential of SED
modeling to provide inputs to theoretical work on evolutionary tracks.
The power law indices for disk and envelope accretion rates are about the same which implies the fraction of
the infalling material lost to the outflow is independent of mass.

Combining evolutionary track results and our SED model results, and taking the indices to 
be the same in the above two power laws, we can obtain 
a value for $f$, the fraction of the $ \dot M_{env}$ eventually arriving on the star 
to be $f \approx 10^{-5}/10^{-4.2} = 0.16$. Conversely, the fraction lost to the outflow is
$f^\prime = 1-f = 0.84 $. This is to compared with a value of 15
star-formation and the theoretically indicated value of $f^\prime \cong 1/3 (f \cong 2/3) $ for low-mass star-formation 
(Tomisaka 1998, Shu et al 1999).

\section{Summary and Conclusions}

Combining a large archive of 2-D radiative transfer models 
(Robitaille et al 2006, 2007) and SEDs with wide wavelength coverage 
using the Spitzer GLIMPSE and MIPSGAL surveys, and the MSX and our published millimeter and sub-millimeter surveys,
a carefully chosen subset of 13 HMPO candidates
was modeled. The main findings are: (1) the models fit the data well with
high implied envelope accretion rates 
${\dot{M}_{env}} \approx 10^{-2.5} \msun/yr$, required for massive
star-formation, lending credence to accretion based massive star-formation. 
(2) stars of masses  $> 20 \msun$ may form by accretion. We also find a possible 
mass dependence of the accretion rates as a power law and determine the fraction
of the envelope accretion lost to the the outflow jets, both of which are subject to large
uncertainties. These results can be improved by extension to larger samples using archival data
and better PSF fitting to overcome saturation.

\acknowledgments 
FMF thanks the organizers of the NSF funded (Award \#0243915) REU Summer Intern Program at the
Harvard-Smithsonian Center for Astrophysics.
This work used NASA's ADS Bibliographic Services, the SIMBAD database 
and the NASA/IPAC Infrared Science Archive.


\clearpage
\begin{table}
\caption{The Sample of HMPO Candidates}

    \begin{tabular}{ccccc}
      \hline
      \hline
      Source & R.A. & Decl. & $d_{f}$ & $d_{n}$\\
      IRAS& J2000.0 & J2000.0 & kpc & kpc\\
      \hline
      18090--1832 & 18 12 01.9 & --18 31 56 & 10.0 & 6.6\\
      18247--1147 & 18 27 31.6 & --11 45 55 & (9.3) & 6.7\\
      18264--1152 & 18 29 14.7 & --11 50 23 & 12.4 & 3.5\\
      18372--0541 & 18 39 55.9 & --05 38 45 & 13.4 & (1.8)\\
      18431--0312 & 18 45 45.9 & --03 09 25 & 8.2 & 6.7\\
      18440--0148 & 18 46 36.6 & --01 45 23 & 8.3 &\\
      18472--0022 & 18 49 52.5 & --00 18 57 & 11.1 & (3.2)\\
      18521+0134 & 18 54 40.7 & +01 38 07 & 9.0 & 5.0\\
      18553+0414 & 18 57 53.4 & +04 18 18 & 12.9 & (0.6)\\
      19035+0641 & 19 06 01.6 & +06 46 36 & 2.2 &\\
      19074+0752 & 19 09 53.6 & +07 57 15 & 8.7 &\\
      19217+1651 & 19 23 58.8 & +16 57 41 & 10.5 &\\
      19266+1745 & 19 28 55.6 & +17 51 60 & 10.0 & (0.3)\\
      \hline \\
    \end{tabular}

\end{table}
\clearpage

\clearpage

\begin{table*}
 \caption{Derived Parameters}
\scriptsize
    \begin{tabular}{c|c|c|c|c|cc|cc|cc|cc|ccc}
      \hline
      \hline
      source & far/ & $\Delta N$ & $\#$ of &
$\overline{\chi_{best}^2}$\tablenotemark{1} & \multicolumn{2}{c}{$M_{\ast} [\msun]$} &
\multicolumn{2}{c}{$\dot{M_{env}} [log\msun/yr]$} &
\multicolumn{2}{c}{$L_{\ast} [log\lsun]$} & \multicolumn{2}{c}{$T_{\ast} [log K]$} &
\multicolumn{3}{c}{$R_{in} [AU]$}\\
      & near & & models & & avg & $\Delta$ & avg & $\Delta$ & avg & $\Delta$ & avg & $\Delta$ & min & best & max\\
      \hline
      18090--1832 & N & 3 & 041 & 2.39 & 14.3 & 3.0 & -2.42 & 0.57 & 3.83 & 0.37 & 3.69 & 0.29 & 03.3 & 07.9 & 57.4\\
      & F & 3 & 026 & 2.72 & 17.8 & 3.3 & -2.26 & 0.52 & 4.15 & 0.55 & 3.72 & 0.33 & 04.3 & 06.2 & 23.8\\
      18247--1147 & N & 3 & 022 & 1.12 & 18.0 & 3.0 & -2.18 & 0.33 & 4.15 & 0.34 & 3.79 & 0.46 & 06.2 & 09.7 & 22.0\\
      18264--1152 & N & 3 & 013 & 8.01 & 12.5 & 2.1 & -2.88 & 1.12 & 3.83 & 0.51 & 3.71 & 0.36 & 03.3 & 04.3 & 13.6\\
      & F & 5 & 004 & 10.00 & 27.4 & 0.0 & -2.48 & 0.00 & 5.16 & 0.00 & 4.55 & 0.00 & 69.5 & 69.5 & 69.5\\
      18372--0541 & F & 5 & 007 & 2.51 & 18.9 & 4.0 & -2.44 & 0.44 & 4.80 & 0.43 & 4.41 & 0.39 & 13.9 & 27.3 & 69.5\\
      18431--0312 & N & 3 & 017 & 2.62 & 11.3 & 1.0 & -2.53 & 0.22 & 3.90 & 0.43 & 4.07 & 0.34 & 04.1 & 08.4 & 22.3\\
      & F & 3 & 011 & 2.85 & 13.1 & 2.3 & -2.41 & 0.29 & 3.96 & 0.47 & 4.05 & 0.59 & 03.8 & 07.4 & 27.4\\
      18440--0148 & E & 3 & 070 & 1.07 & 15.8 & 2.6 & -2.38 & 0.35 & 4.37 & 0.64 & 4.33 & 0.49 & 05.8 & 26.1 & 86.6\\
      18472--0022 & F & 5 & 005 & 3.77 & 16.0 & 2.1 & -2.51 & 0.16 & 4.77 & 0.04 & 4.47 & 0.05 & 27.3 & 27.3 & 50.4\\
      18521+0134 & N & 3 & 063 & 0.96 & 14.2 & 3.2 & -2.41 & 0.51 & 3.83 & 0.43 & 3.74 & 0.62 & 03.3 & 07.4 & 85.6\\
      & F & 3 & 021 & 2.54 & 18.6 & 3.0 & -2.24 & 0.36 & 4.27 & 0.55 & 3.81 & 0.39 & 06.2 & 08.3 & 23.8\\
      18553+0414 & F & 5 & 003 & 2.91 & 17.0 & 2.6 & -2.48 & 0.21 & 4.74 & 0.15 & 4.39 & 0.39 & 13.9 & 27.3 & 50.4\\
      19035+0641 & E & 3 & 136 & 1.13 & 10.3 & 1.8 & -2.66 & 0.54 & 3.63 & 0.75 & 3.95 & 0.76 & 01.8 & 22.3 & 85.6\\
      19074+0752 & E & 3 & 013 & 3.41 & 17.4 & 3.4 & -2.16 & 0.31 & 4.30 & 0.77 & 4.00 & 0.81 & 06.4 & 11.0 & 27.3\\
      19217+1651 & E & 5 & 008 & 7.86 & 21.8 & 3.9 & -2.34 & 0.55 & 4.83 & 0.52 & 4.30 & 0.58 & 13.9 & 27.3 & 69.5\\
      19266+1745 & F & 3 & 010 & 7.15 & 14.8 & 2.2 & -2.35 & 0.67 & 4.10 & 1.14 & 3.88 & 0.90 & 04.6 & 09.7 & 27.3\\
      \hline
      \multicolumn{16}{c}{1. Refers to the $\chi_{best}^2$ divided by the number of flux data points (excluding the upper limit points).}
    \end{tabular}
\end{table*}

\tiny

\begin{table}
\caption{Fluxes of the HMPO Candidates}
\setlength{\tabcolsep}{3pt}
\tiny
  \begin{tabular}{crrrrrrrrrrrrrr}
  \multicolumn{15}{c}{Table 3: Fluxes of the HMPO Candidates}\\
      \hline
      \hline
      Source & IRAC & IRAC & IRAC & IRAC & MIPS & MIPS & MSX & MSX & MSX & MSX & SCUBA & SCUBA & IRAM & Aperture\\
      IRAS & $3.6\ \mu m$ & $4.5\ \mu m$ & $5.8\ \mu m$ & $8.0\ \mu m$ & $24\ \mu m$ & $70\ \mu m$ & $8.3\ \mu m$ & $12.1\ \mu m$ & $14.7\
\mu m$ & $21.3\ \mu m$ & $450\ \mu m$ & $850\ \mu m$ & $1.2\ mm$ & (sub)-mm\\
      & (mJy) & (mJy) & (mJy) & (mJy) & (Jy) & (Jy) & (Jy) & (Jy) & (Jy) & (Jy) & (Jy) & (Jy) & (Jy) & ($''$)\\
      \hline
      18090\--1832 & ... & ... & 835.3 & 797.7 & 5.1 & 56.4 & 0.80 & 1.65 & 3.05 & 5.61 & 8.6 & 4.5 & 0.6 & 12\\
      & & & $\pm$ 208.8 & $\pm$ 191.9 & $\pm$ 1.3 & $\pm$ 14.1 & $\pm$ 0.20 & $\pm$ 0.41 & $\pm$ 0.76 & $\pm$ 1.40 & $\pm$ 2.2 & $\pm$
1.1 & $\pm$ 0.2 &\\
      18247\--1147 & 89.2 & 203.7 & 574.4 & ... & ... & 156.7 & 2.49 & 9.14 & 16.57 & 51.92 & 32.1 & 6.9 & 1.9 & 13\\
      & $\pm$ 25.4 & $\pm$ 50.9 & $\pm$ 143.4 & & & $\pm$ 39.2 & 90 \% & 90 \% & 90 \% & 90 \% & $\pm$ 8.0 & $\pm$ 1.7 & $\pm$ 0.5 &\\
      18264\--1152 & 58.9 & 357.8 & 770.1 & 658.6 & 4.4 & 183.7 & 0.61 & 0.49 & 0.73 & 3.43 & 106.6 & 20.0 & 7.9 & 15\\
      & 90 \% & 90 \% & 90 \% & 90 \% & $\pm$ 1.1 & $\pm$ 45.9 & $\pm$ 0.15 & $\pm$ 0.49 & $\pm$ 0.18 & $\pm$ 0.86 & $\pm$ 26.7 & $\pm$
5.0 & $\pm$ 2.0 &\\
      18372\--0541 & 27.8 & 192.1 & 491.6 & 698.5 & ... & 138.7 & 1.05 & 2.46 & 4.65 & 11.54 & 12.9 & 4.3 & 1.4 & 24\\
      & $\pm$ 7.0 & $\pm$ 48.0 & $\pm$ 122.9 & $\pm$ 174.6 & & $\pm$ 34.7 & 90 \% & 90 \% & 90 \% & 90 \% & $\pm$ 3.2 & $\pm$ 1.1 &
$\pm$ 0.4 &\\
      18431\--0312 & 97.7 & 190.2 & 290.4 & ... & 6.3 & 80.5 & 2.31 & 2.49 & 1.64 & 5.07 & 17.1 & 3.1 & 1.1 & 31\\
      & $\pm$ 24.4 & $\pm$ 47.6 & $\pm$ 72.6 & & $\pm$ 1.6 & $\pm$ 20.1 & 90 \% & 90 \% & 90 \% & $\pm$ 1.27 & $\pm$ 4.3 & $\pm$ 0.8 &
$\pm$ 0.3 &\\
      18440\--0148 & ... & ... & 681.0 & 561.1 & ... & 145.4 & 0.94 & 1.43 & 2.80 & 14.68 & 19.8 & 2.6 & 0.5 & 13\\
      & & & $\pm$ 170.3 & $\pm$ 140.3 & & $\pm$ 36.3 & $\pm$ 0.23 & $\pm$ 0.36 & $\pm$ 0.70 & $\pm$ 3.67 & $\pm$ 5.0 & $\pm$ 0.7 & $\pm$
0.1 &\\
      18472\--0022 & 182.1 & 315.9 & 497.0 & 667.2 & ... & 130.0 & 1.64 & 3.44 & 4.25 & 16.66 & 45.3 & 7.4 & 2.7 & 30\\
      & $\pm$ 45.5 & $\pm$ 79.0 & $\pm$ 124.3 & $\pm$ 166.8 & & $\pm$ 32.5 & $\pm$ 0.41 & $\pm$ 0.86 & $\pm$ 1.06 & $\pm$ 4.41 & $\pm$
11.3 & $\pm$ 1.9 & $\pm$ 0.7 &\\
      18521+0134 & 69.9 & ... & 1175.0 & 1279.0 & ... & 96.8 & 1.49 & 2.19 & 3.35 & 7.21 & 21.6 & 3.3 & 1.0 & 16\\
      & $\pm$ 17.5 & & $\pm$ 293.8 & $\pm$ 319.8 & & $\pm$ 24.2 & $\pm$ 0.37 & $\pm$ 0.55 & $\pm$ 0.84 & $\pm$ 1.80 & $\pm$ 5.4 & $\pm$
0.8 & $\pm$ 0.3 &\\
      18553+0414 & 76.5 & 445.8 & 1440.0 & ... & ... & 122.2 & 2.75 & 5.65 & 8.62 & 15.92 & 23.3 & 4.3 & 1.9 & 15\\
      & $\pm$ 19.1 & $\pm$ 111.4 & 90 \% & & & $\pm$ 30.6 & 90 \% & 90 \% & 90 \% & 90 \% & $\pm$ 5.8 & $\pm$ 1.1 & $\pm$ 0.5 &\\
      19035+0641 & ... & 304.6 & 958.0 & 1266.0 & ... & 265.6 & 0.98 & 2.29 & 4.94 & 29.07 & 72.8 & 10.6 & 2.6 & 32\\
      & & $\pm$ 76.2 & $\pm$ 239.5 & $\pm$ 316.5 & & $\pm$ 66.4 & $\pm$ 0.25 & $\pm$ 0.57 & $\pm$ 1.24 & $\pm$ 7.27 & $\pm$ 18.2 & $\pm$
2.7 & $\pm$ 0.7 &\\
      19074+0752 & ... & ... & 235.4 & 439.1 & ... & 119.2 & 2.33 & 4.64 & 5.37 & 23.51 & 26.2 & 6.8 & 0.8 & 42\\
      & & & $\pm$ 58.9 & $\pm$ 109.8 & & $\pm$ 29.8 & 90 \% & 90 \% & 90 \% & 90 \% & $\pm$ 6.5 & $\pm$ 1.7 & $\pm$ 0.2 &\\
      19217+1651 & 104.3 & ... & 1370.0 & 1380.0 & ... & 191.1 & 1.01 & 1.39 & 3.62 & 12.55 & 68.5 & 6.9 & 2.6 & 16\\
      & $\pm$ 26.1 & & $\pm$ 342.5 & $\pm$ 345.0 & & $\pm$ 47.8 & $\pm$ 0.25 & $\pm$ 0.35 & $\pm$ 0.90 & $\pm$ 3.14 & $\pm$ 17.1 & $\pm$
1.7 & $\pm$ 0.7 &\\
      19266+1745 & ... & 20.1 & 90.2 & 212.8 & ... & 66.0 & 0.90 & 1.28 & 1.01 & 5.46 & 36.4 & 6.0 & 1.3 & 20\\
      & & $\pm$ 5.0 & $\pm$ 22.6 & $\pm$ 53.2 & & $\pm$ 16.5 & $\pm$ 0.22 & $\pm$ 0.32 & $\pm$ 0.25 & $\pm$ 1.37 & $\pm$ 9.1 & $\pm$ 1.5
& $\pm$ 0.3 &\\

\hline
\end{tabular}
\end{table}

\clearpage

\begin{figure}[htp!b]
\includegraphics[width=7.0in]{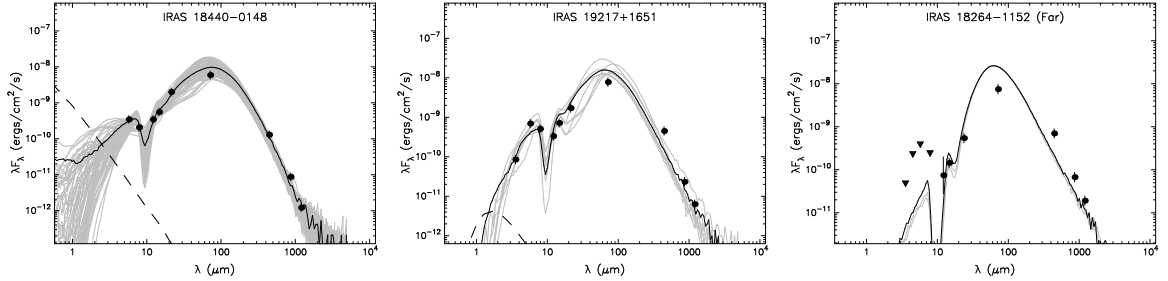} 
\caption{SEDs for the sources outlined in Table 1. (a) The three panels show representative (IRAS 18440-0148 \& IRAS 19217+1651l)
and the worst (IRAS 18264-1152-far) fits. 
The triangles denote upper limits. The solid black line shows the
best-fitting model while the gray lines show all models that also fit
the data reasonably well (${\chi}^2 - {\chi_{best}}^2 < \Delta N$ where
$\Delta N$ = 1, 3 or 5).  Since the apertures used varied between
wavelengths, the fits plotted are synthetic SEDs obtained by
interpolating
between the different apertures.
The dashed line shows the SED of the stellar photosphere in the best-fitting model. 
}
\end{figure}
\clearpage
\setcounter{figure}{0}
\begin{figure}
\includegraphics[width=6.3in]{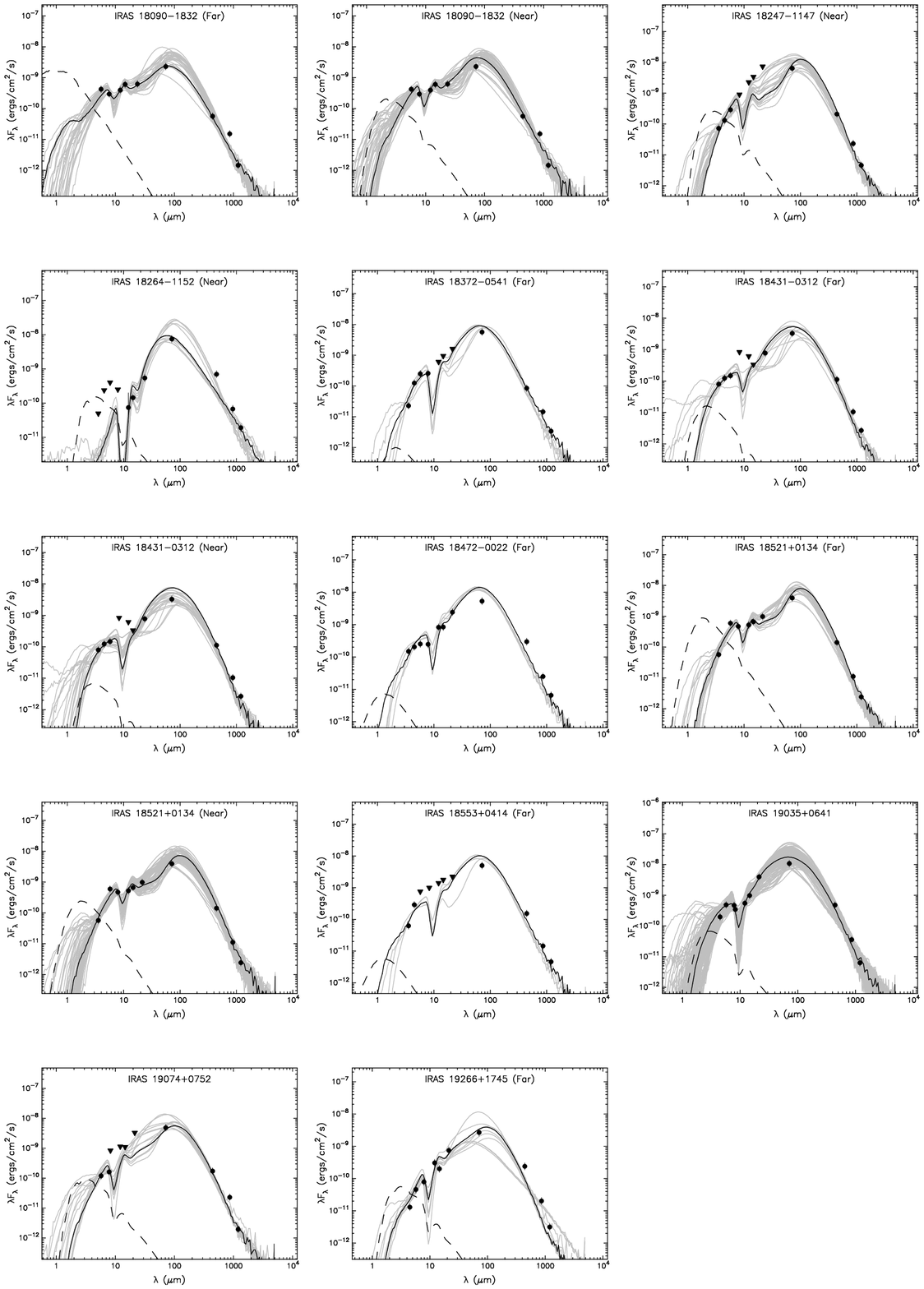}
\caption {(b) same as (a), for the full set of objects}
\end{figure}
\vspace{-0.5cm}

\clearpage
\begin{figure}
\includegraphics[angle=-90, width=6in]{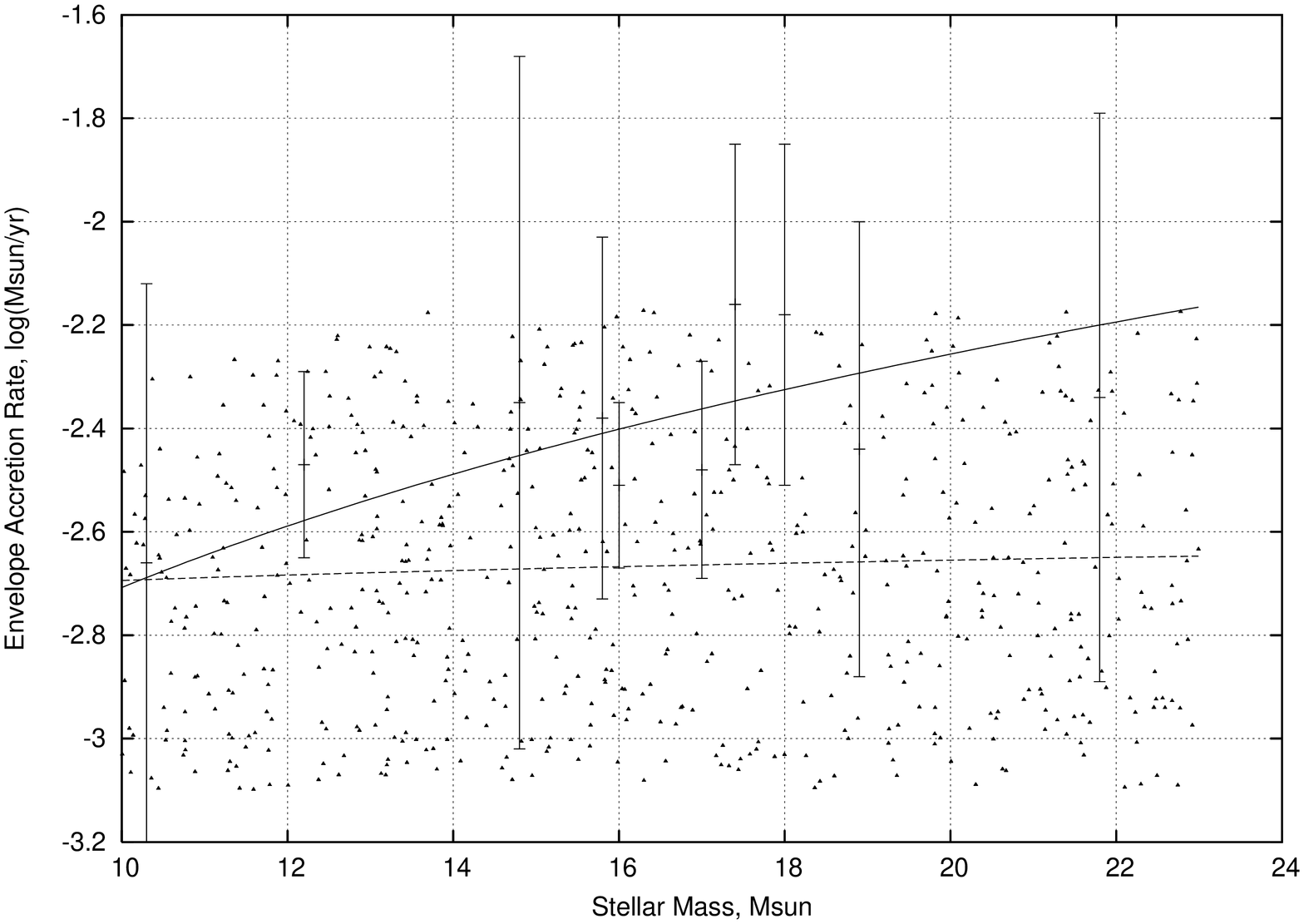}
\caption{Plot of envelope accretion rate $\dot{M_{env}}$ versus stellar mass
$M_{\ast}$ for the sources. The solid line is the best-fit power law to the
data. This result is not biased by the model grid distribution, also shown, with the dashed line being 
its best fit.} 
\end{figure}
\end{document}